\documentclass[useAMS,usenatbib, usegraphicx]{mn2e}
\usepackage{epsfig}
\usepackage{fancyhdr}
\usepackage{makeidx}
\usepackage{titletoc}
\usepackage{lscape}
\usepackage{hyperref}
\usepackage{wasysym}
\usepackage{wrapfig}
\usepackage{longtable}
\usepackage{aas_macros}
\usepackage{txfonts}
\usepackage{times}

\begin{document}

\title[The Alpha-element knee of the Sagittarius Stream]{The Alpha-element knee of the Sagittarius Stream}
\author[T.J.L. de Boer, V. Belokurov, T.C. Beers and Y.S. Lee]{T.J.L. de Boer$^{1}$\thanks{E-mail:
tdeboer@ast.cam.ac.uk}, V. Belokurov$^{1}$, T.C. Beers$^{2,3}$ and Y.S. Lee$^{4}$\\
$^{1}$ Institute of Astronomy, University of Cambridge, Madingley Road, Cambridge CB3 0HA, UK \\
$^{2}$ Department of Physics \& Astronomy and JINA (Joint Institute for Nuclear Astrophysics), Michigan State University, East Lansing, MI 48824, USA \\
$^{3}$ National Optical Astronomy Observatory, 950 N. Cherry Avenue, Tucson, AZ 85719, USA\\
$^{4}$ Department of Astronomy, New Mexico State University, Las Cruces, NM 88003, USA}
\date{Received ...; accepted ...}

\pagerange{\pageref{firstpage}--\pageref{lastpage}} \pubyear{2014}

\maketitle

\begin{abstract}
We employ abundances from the Sloan Digital Sky Survey~(SDSS) and the Sloan Extension for Galactic Understanding and Exploration~(SEGUE) to study the $\alpha$-element distribution of the stellar members of the Sagittarius stream. To test the reliability of SDSS/SEGUE abundances for the study of Sagittarius, we select high-likelihood samples tracing the different components of the Milky Way, and recover known literature $\alpha$-element distributions. \\
Using selection criteria based on the spatial position, radial velocity, distance and colours of individual stars, we obtain a robust sample of Sagittarius-stream stars. The $\alpha$-element distribution of the Sagittarius stream forms a narrow sequence at intermediate metallicities with a clear turn-down, consistent with the presence of an $\alpha$-element ''knee". This is the first time that the $\alpha$-element knee of the Sagittarius dwarf galaxy has been detected. Fitting a toy model to our data, we determine that the $\alpha$-knee in Sagittarius takes place at [Fe/H]=$-$1.27$\pm$0.05, only slightly less metal-poor than the knee in the Milky Way. This indicates that a small number of Sagittarius-like galaxies could have contributed significantly to the build-up of the Milky Way's stellar halo system at ancient times.
\end{abstract}


\label{firstpage}

\section{Introduction}
\label{introduction}
The Galactic stellar halo system is expected to have been formed during the initial stages of Milky Way~(MW) formation, through the accretion of smaller stellar systems, such as dwarf galaxies~\citep[e.g.,][]{Kauffmann93, Navarro97,Bullock05}. Spectroscopic studies of the MW halo have shown that it is dominated by metal-poor stars~([Fe/H]$<$$-$1.0) with [$\alpha$/Fe]$\approx$+0.4, consistent with chemical enrichment solely by massive stars, before Supernovae~(SNe) of type Ia began to contribute significantly to the ISM~\citep[e.g.,][]{McWilliam97, Matteucci03, Venn04}. Therefore, if the halo system was formed through the merger of dwarf galaxies, their abundance patterns should be consistent with that of the MW halo. High-Resolution~(HR) spectroscopic studies of "classical" Local Group~(LG) dSphs have shown that their current $\alpha$-abundances are inconsistent with the abundance pattern of the old MW halo, ruling out halo formation through the accretion of present-day dwarf galaxies. However, some LG dSphs contain a significant fraction of old, metal-poor stars with alpha-abundances consistent with the MW halo system, indicating that early accretion of these systems could be sufficient to form parts of the stellar MW halo~\citep[e.g.,][]{Tolstoy09, Kirby11, deBoer2012A}. \\
An important missing piece in this puzzle is the range of masses of dwarf galaxies that exhibit abundance patterns consistent with the MW halo, and the timescale within which they would need to be accreted. The metallicity of the turn-over in $\alpha$-element abundances~(the so-called $\alpha$-knee) is linked to the Star Formation Rate during the early stage of star formation in a galaxy and therefore also depends on the total mass of the system. Therefore, early accretion of a large number of low mass dwarf galaxies~(M$_{dyn}$$\le$10$^{8}$ M$_{\odot}$) could assemble the most metal-poor~([Fe/H]$<$$-$1.5) parts of the MW halo system, but cannot easily reproduce the more metal-rich~([Fe/H]$>$$-$1.5), high [$\alpha$/Fe] stars in the old MW halo~\citep[e.g.,][]{Tolstoy09}. To assemble the old, metal-rich MW halo, we need to invoke the accretion of more massive dwarf galaxies, for which the $\alpha$-knee takes place at higher metallicity. Furthermore, given their large mass it is also possible to form the entire old MW halo through the accretion of only a relatively small number of massive dwarf galaxies at early times. \\
The Sagittarius dwarf spheroidal galaxy~(Sgr dSph) and associated stellar streams provide valuable tools for the study of galaxy formation and evolution. The Sgr dSph was discovered by~\citet{Ibata95} as the nearest dwarf galaxy to the Milky Way~(MW), at a distance of~$\approx$25 kpc. Subsequent studies showed that Sgr was severely influenced by Galactic tides, pulling large numbers of stars from the core to form stellar streams that wrap around the Galaxy at least once~\citep[e.g.,][]{Johnston95,Lynden-Bell95, Ibata01, Majewski03, Belokurov06}. With an estimated dark matter mass of $\approx$10$^{9}$ M$_{\odot}$, the Sgr dSph is the third most massive surviving satellite of the MW~\citep{Niederste-Ostholt10}. Therefore, by studying the abundance pattern of Sgr, we can determine if it is possible to form the MW halo through the merger of Sgr-like dwarf galaxies at early times. \\
Previous studies of $\alpha$-element abundances in Sgr have focussed on the main body of the dwarf galaxy, since it is easier to select member stars in the relatively crowded central regions. These studies show that Sgr displays a sequence of [$\alpha$/Fe] ratios ranging from [$\alpha$/Fe]=+0.2 at [Fe/H]=$-$1.0 down to [$\alpha$/Fe]=$-$0.2 at [Fe/H]=0.0~\citep[e.g.,][]{Monaco05, Sbordone07, Carretta102, McWilliam13}. Unfortunately, the HR results do not determine the $\alpha$-element distribution at low metallicities~([Fe/H]$<$$-$1.5), since the Sgr main body is dominated by metal-rich stars. To obtain a robust determination of the Sgr $\alpha$-element distribution at lower metallicities, we need to study the Sgr streams, which are dominated by more metal-poor stars. However, due to the large area subtended on the sky, it is challenging to obtain large samples of spectroscopic abundances of stars in the Sgr stream. \\
In this paper, we make use of data from the Sloan Digital Sky Survey~(SDSS) Data Release 10~\citep{Ahn14} and the Sloan Extension for Galactic Understanding and Exploration~(SEGUE), to study the $\alpha$-element distribution of the Sgr stream. The SEGUE survey has obtained medium-resolution spectra of stars covering a large portion of the sky, including many fields consistent with the Sgr stream~\citep{Yanny091}. Estimates of the [$\alpha$/Fe] ratio~(defined as [$\alpha$/Fe]=0.5$\times$[Mg/Fe] + 0.3$\times$[Ti/Fe]+0.1$\times$[Ca/Fe]+0.1$\times$[Si/Fe]) for individual stars have been determined through detailed synthetic spectrum fitting using the SEGUE Stellar Parameter Pipeline~\citep[SSPP,][]{SEGUE1,SEGUE2,SEGUE3,SEGUE4,SEGUE5}. We first determine the sensitivity of the SDSS/SEGUE [$\alpha$/Fe] ratios by comparing them to results from HR spectroscopic studies of different MW components. Subsequently, we select a sample of high-confidence Sgr-stream stars by making use of spatial position and radial velocity cuts. Finally, we compare the $\alpha$-element distribution of the Sgr stream to that of the MW and other LG dSphs, to explore the range of masses of dwarf galaxies that could have contributed to the formation of the MW stellar halo system.
\begin{figure*}
\centering
\includegraphics[angle=0, width=0.49\textwidth]{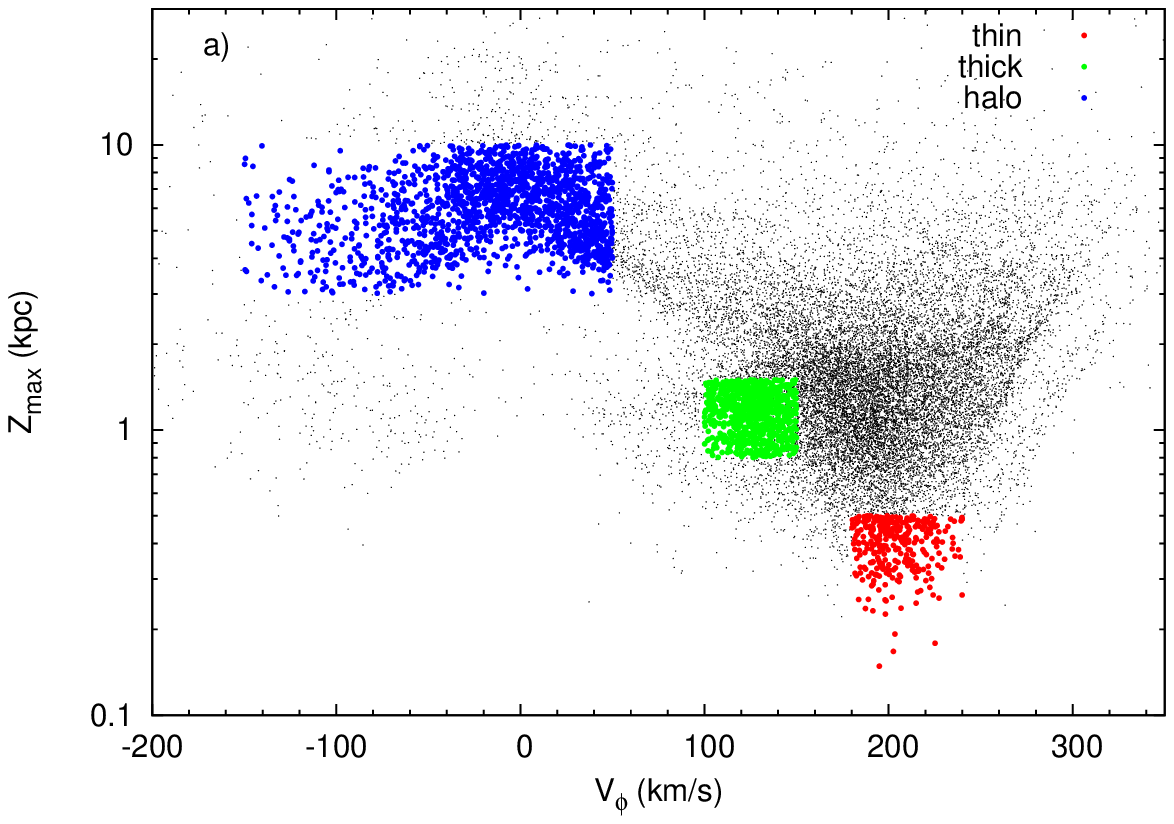}
\includegraphics[angle=0, width=0.49\textwidth]{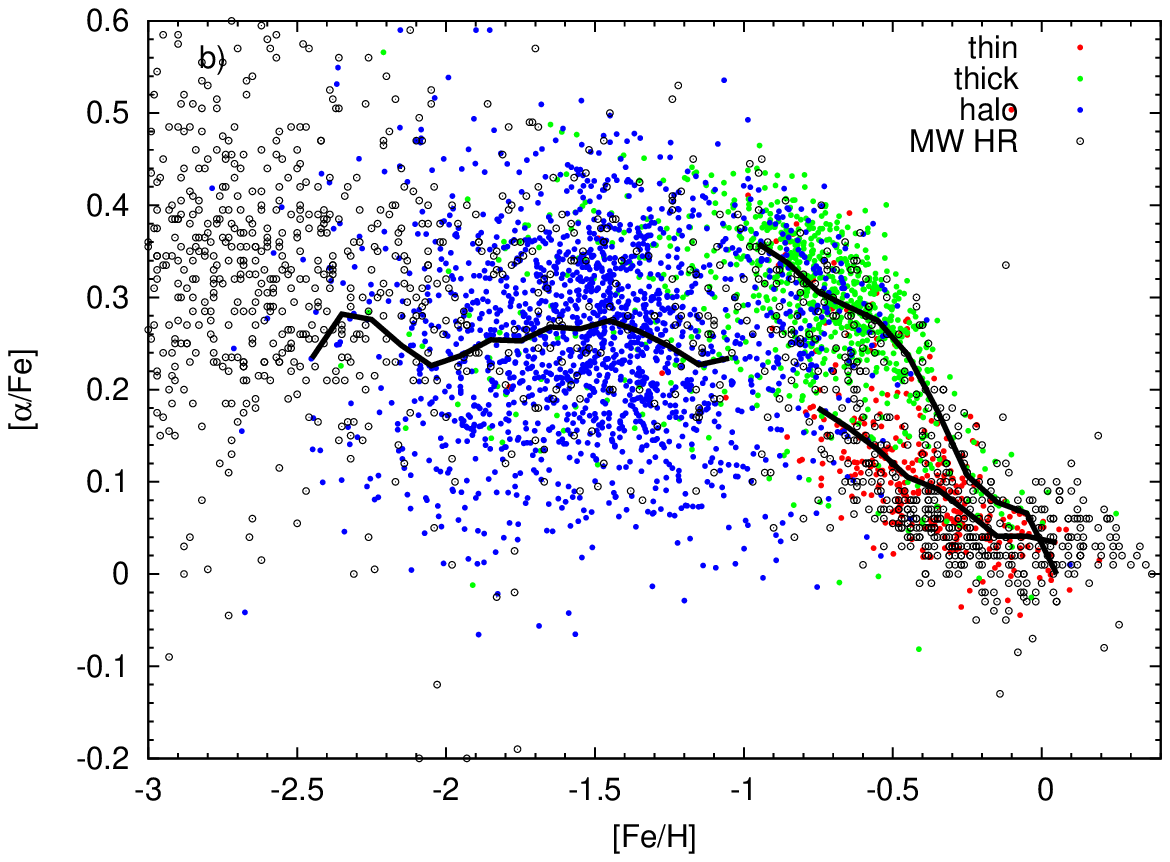}
\caption{\textbf{Left:} The distribution of V$_{\phi}$ vs Z$_{max}$ as determined from the orbit fitting of individual stars. Coloured points indicate the selections adopted for each MW component, thin disc~(red), thick disc~(green) and halo~(blue). \textbf{Right:} The abundance of [$\alpha$/Fe] as a function of [Fe/H] for the different components of the MW selected from the SEGUE survey, in comparison to HR spectroscopic samples of MW stars~\citep{Venn04} . Solid lines indicate the median [$\alpha$/Fe] abundance at each metallicity for the different components. \label{MWcomparison}}
\end{figure*}

\section{SEGUE alpha-element abundance in the Milky Way}
\label{mwcomparison}
The [$\alpha$/Fe] abundances of stars in the SDSS/SEGUE survey are determined from spectra with a much lower resolution than adopted in classical abundance analyses. Therefore, it is important to determine the reliability of the abundances across the range of [$\alpha$/Fe] we are interested in. For a detailed analysis of the validity of SDSS/SEGUE $\alpha$-element abundances, we refer the reader to~\citet{SEGUE5}. Here, we test the SDSS/SEGUE abundances by selecting a high-likelihood sample of MW stars, and determining the [$\alpha$/Fe] distribution of different MW components. These distributions can then be compared to results from classical HR studies to determine if we can reproduce the MW $\alpha$-element abundances over different metallicity and [$\alpha$/Fe] ranges. \\
We first separate our MW sample into different components, using orbital parameters derived from the full 6d phase-space information of individual stars. We obtain the positions and radial velocities of each star from the SDSS \textit{phototag} and \textit{sppparams} tables. Furthermore, the proper motions are available from the SDSS \textit{propermotions} table, determined by comparing the positions of stars in SDSS to those from recalibrated positions from the USNO-B catalog~\citep{Munn04, Munn08}. To determine distances to stars in our sample, we use the relations for photometric parallax from~\citet{Ivezic08} to obtain the distance modulus for dwarfs with SDSS colours. To obtain reliable distances, we adopt the same colour selections as given in~\citet{Ivezic08}, and limit our sample to dwarfs with log $g$$>$4 and 4500$\le$T$_{\rm eff}$$\le$6500 within10 kpc of the Sun. Furthermore, we also require that stars in our sample have S/N$\ge$50 and low extinction~(A$_{V}$$\le$0.3). \\
Using the 6d phase-space information, we obtain the orbital parameters of each star using the potential described in~\citet{Fellhauer00,Fellhauer06}. The stars are integrated forward in a realistic MW potential to determine their minimum and maximum radial extent, maximum height above the MW plane, orbital eccentricity and velocity components. The parameters for the potential for thin-disc, thick-disc and halo are the same as adopted in~\citet{Fellhauer06}. \\
Combining the maximum height above the MW plane that a star can reach~(Z$_{max}$) with the velocity V$_{\phi}$, we obtain a useful metric for separating the different components of the MW. Figure~\ref{MWcomparison}a shows the V$_{\phi}$ vs Z$_{max}$ distribution of stars in our MW sample, along with the selections for the different MW components. Thin-disc stars are chosen to have velocities close to those of the Sun and low height above the plane, while thick-disc stars have lower velocities and higher maximum extension above the plane~\citep{Venn04, Ruchti10}. Finally, to unambiguously select the halo we consider only stars with velocities below V$_{\phi}$=50 km/s and a maximum height above the plane of 3 kpc or greater. \\
Figure~\ref{MWcomparison}b shows the [$\alpha$/Fe] distribution of the different MW components, as a function of [Fe/H]. Solid lines indicate the median [$\alpha$/Fe] abundance for each component at different metallicities. The three components are clearly separated in Figure~\ref{MWcomparison}b, and match the distribution of MW components determined from HR spectroscopic surveys~\citep{Venn04}. Thin-disc stars are found at high metallicities and display low [$\alpha$/Fe] abundances. The metallicities of thick-disc stars partially overlap with the thin-disc, but the two components are clearly separated in [$\alpha$/Fe], similar to~\citet{Venn04}. Our sample of halo stars consists mostly of metal-poor stars with overall high $\alpha$-element abundances, consistent with literature results. Furthermore, the median [$\alpha$/Fe] of the halo decreases for increasing metallicity, consistent with the presence of a metal-rich, low $\alpha$-element halo~\citep{Nissen10}. The spread of [$\alpha$/Fe] of the halo sample is larger than that of HR samples, with a considerable number of halo stars showing [$\alpha$/Fe]$<$+0.2.  \\
In conclusion, both the mean [$\alpha$/Fe] ratios and dispersion of SDSS/SEGUE stars are reliable for stars with~[Fe/H]$>$$-$1.5. For more metal-poor stars, we will only trust the median of the [$\alpha$/Fe] distribution and disregard the width.
\begin{figure*}
\centering
\includegraphics[angle=0, width=0.46\textwidth]{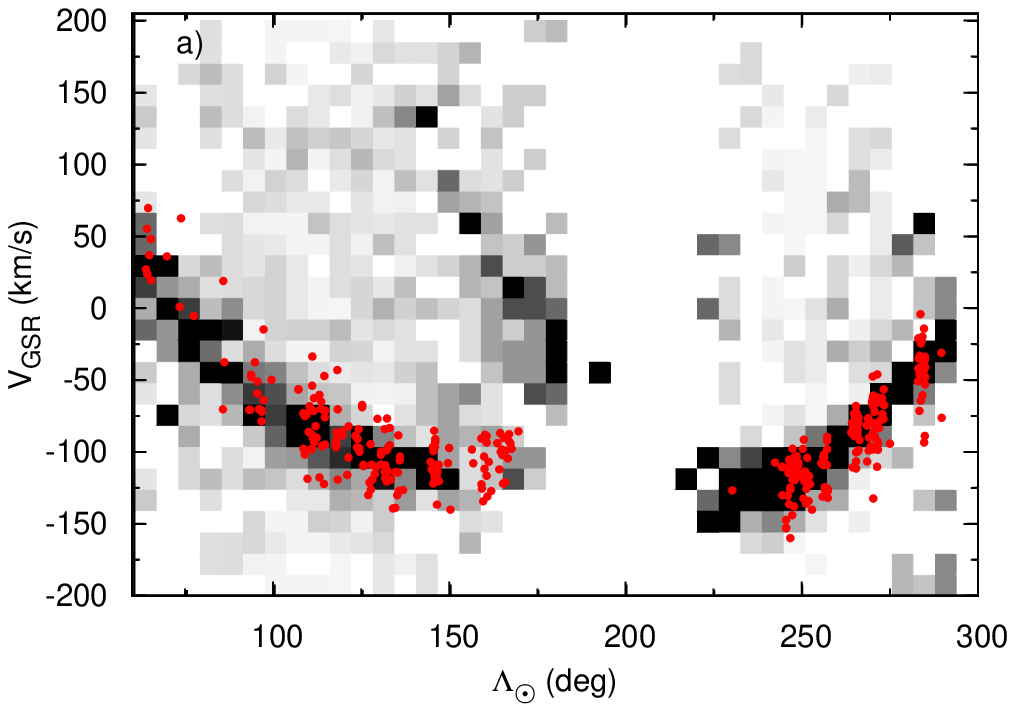}
\includegraphics[angle=0, width=0.53\textwidth]{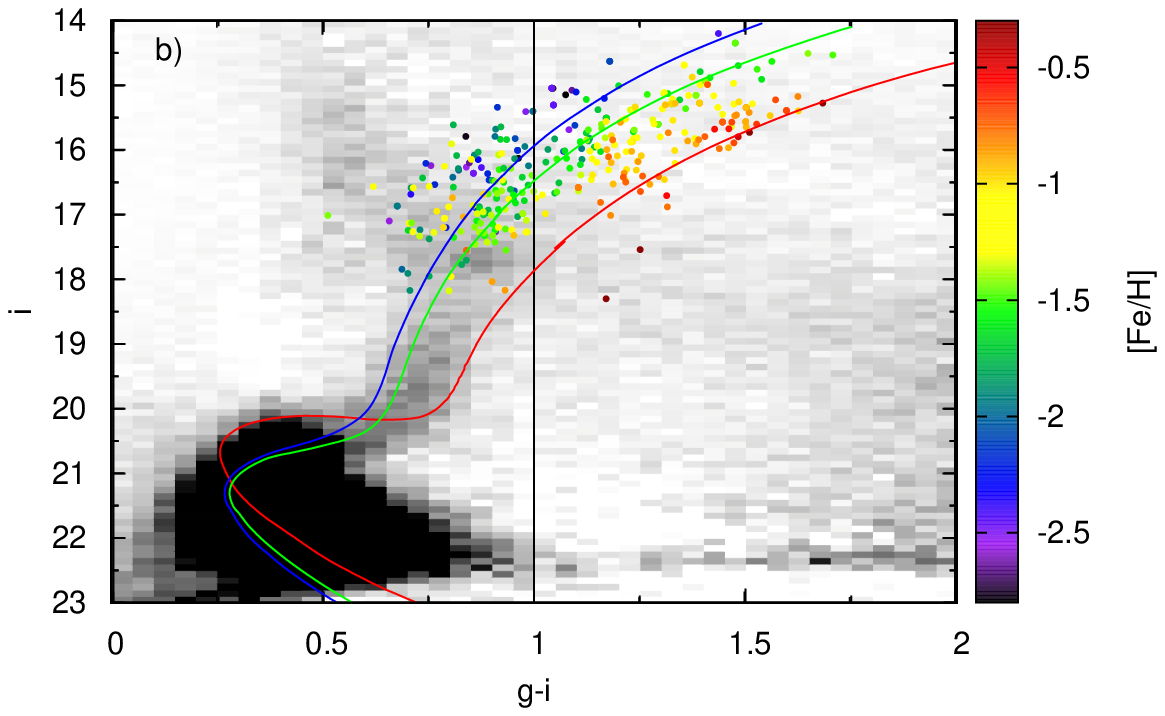}
\caption{\textbf{Left:} Our Sgr sample~(red points) in the $\Lambda$$_{\odot}$ vs Galactocentric line-of-sight velocities V$_{GSR}$ plane. The density of Sgr stars from~\citet{Belokurov14} is also shown for comparison, as the greyscale Hess diagram. \textbf{Right:} The photometric, MW-corrected CMD of Sgr, overlaid with stars from our sample and isochrones representative of Sgr populations~(blue: [Fe/H]=$-$2.2, 14 Gyr, green: [Fe/H]=$-$1.5, 12 Gyr,red: [Fe/H]=$-$0.5, 5 Gyr). The colour of the spectroscopic points indicates the metallicity [Fe/H] of each star. The black line indicates the additional colour selection applied to the final spectroscopic Sgr sample to exclude MW contamination. \label{sgr_selection}}
\end{figure*}

\section{The alpha-element distribution of the Sagittarius stream}
\label{selection}
\begin{figure*}
\centering
\includegraphics[angle=0, width=0.49\textwidth]{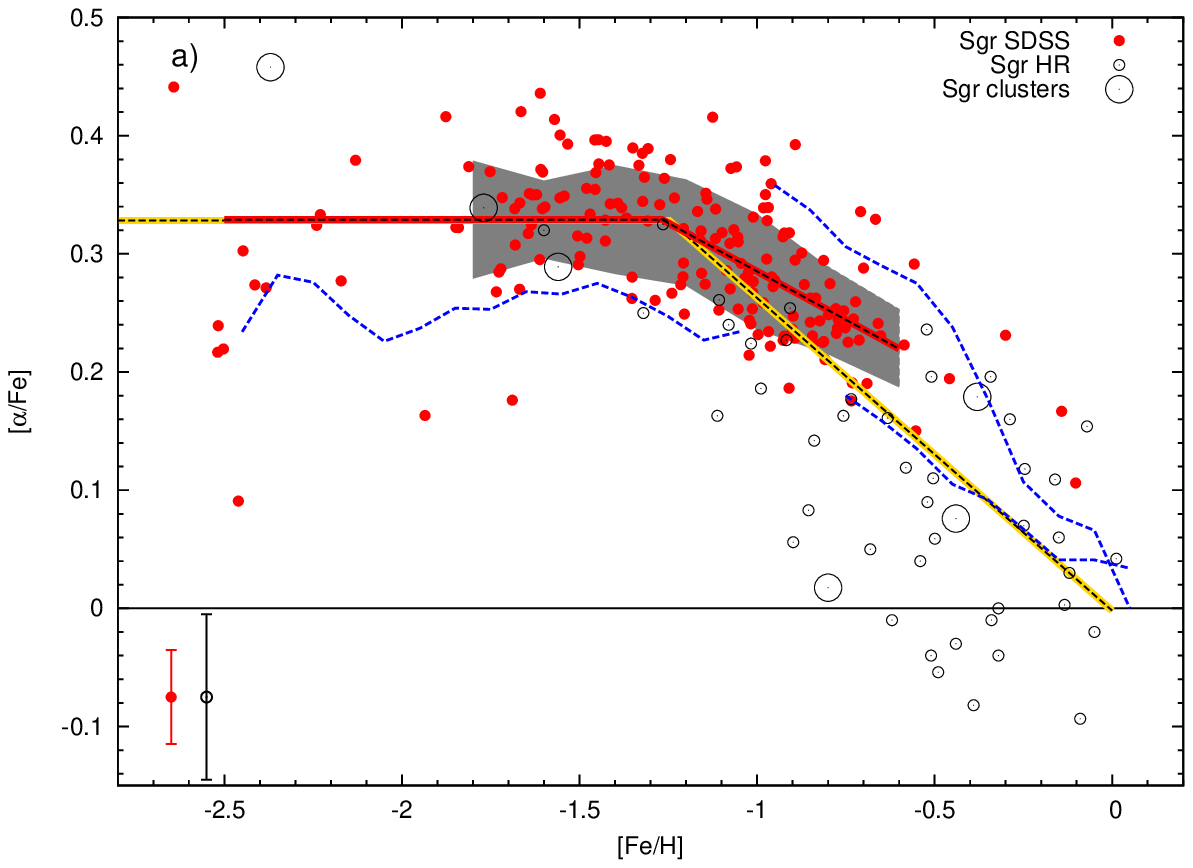}
\includegraphics[angle=0, width=0.49\textwidth]{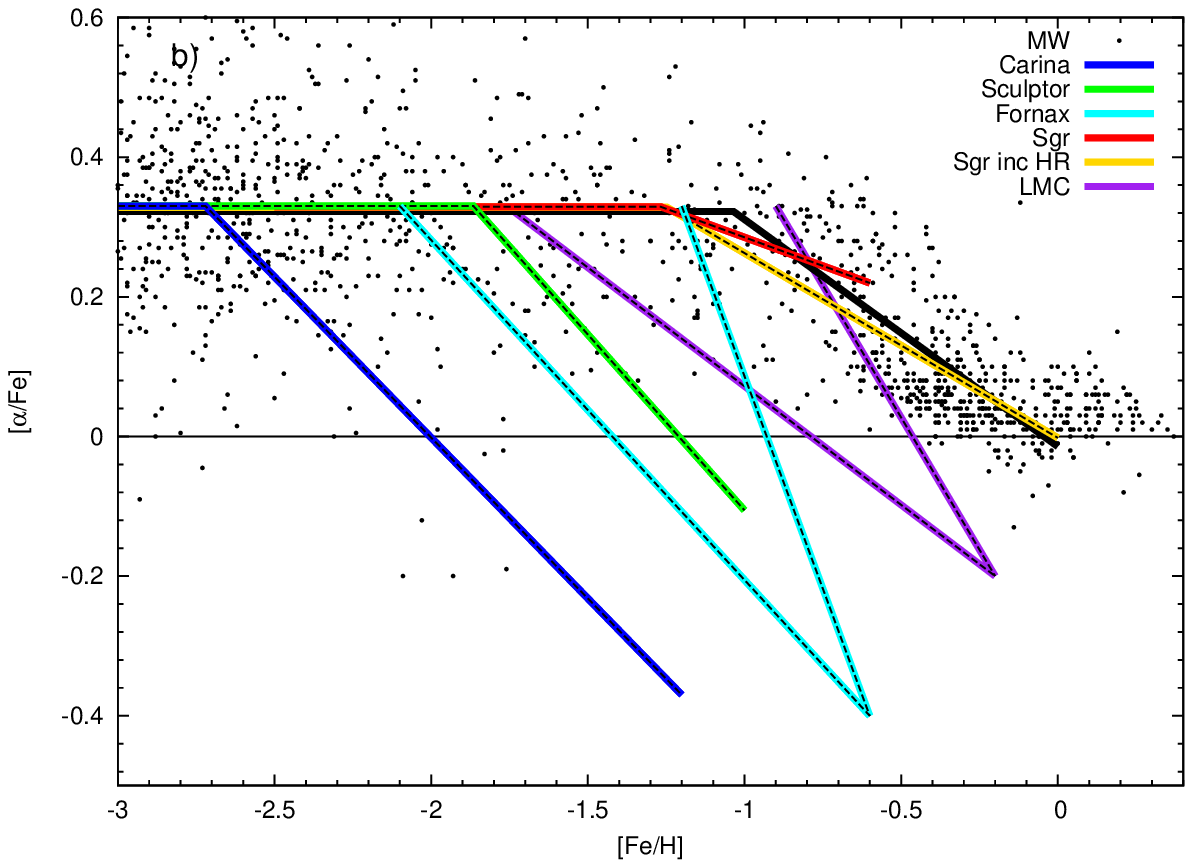}
\caption{\textbf{Left:} The abundances of  [$\alpha$/Fe], as a function of [Fe/H], for our sample of stars belonging to the Sgr stream~(closed red dots), in comparison to HR abundances~(open circles) of the Sgr main body~\citep{SmeckherHane02,Carretta102,McWilliam13} and globular clusters associated to Sgr~\citep{Brown99,Carraro04,Cohen04,Tautvaisiene04,Mottini08}. The solid red line indicates our best-fit toy model for the $\alpha$-element knee, with 1$\sigma$ uncertainties~(solid grey region), while the solid yellow line shows the toy model when fitting the SDSS/SEGUE and HR main-body samples together. Medians for the MW components are also shown as blue dashed lines. \textbf{Right:} The best-fit toy model for the Sgr stream~(red for SDSS/SEGUE sample and yellow for SDSS/SEGUE and HR samples combined) compared to HR abundances of Milky Way stars~(small grey points) and model fits to other LG dSph galaxies~\citep{Venn04, Letarte10,Lemasle12}. For systems with robust knee determinations~(Sgr, MW, Sculptor and Carina) we show the best-fit model, while for other systems~(Fornax and LMC) we show lines tracing the outer envelopes of the abundance distribution. \label{sgr_alphafe}}
\end{figure*}

Having demonstrated the reliability of [$\alpha$/Fe] abundances from SDSS/SEGUE spectra, we now determine the $\alpha$-element distribution of the Sgr stellar stream. The area surveyed by SEGUE includes several fields at locations consistent with the Sgr stream, allowing us to build up a robust sample of stream stars across both the Northern and Southern hemispheres. However, to determine the $\alpha$-element distribution of the stream, we first must correct for effects of contamination from MW halo stars at the same location. \\
To obtain our Sgr sample, we initially select stars based on their position on the sky. For this we make use of the prescriptions of~\citet{Majewski03} to transform equatorial RA and Dec coordinates into heliocentric $\Lambda$$_{\odot}$,B$_{\odot}$ coordinates that are aligned with the Sgr stream. Following the nomenclature of~\citet{Belokurov14}, $\Lambda$$_{\odot}$ increases in the direction of Sgr motion and B$_{\odot}$ points to the North Galactic pole. All stars with -9$<$B$<$9 degrees are assumed to be located within the Sgr plane. Furthermore, we limit our sample to stars with robust stellar parameters and abundances using S/N$\ge$50 as well as spectroscopically confirmed giants using log $g$$\le$3.5, 4300$\le$T$_{\rm eff}$$\le$6000 K. \\
To reduce contamination from MW stars, we exploit the fact that the Sgr stream displays a clear signal in the $\Lambda$$_{\odot}$ vs line-of-sight velocities V$_{GSR}$ plane~\citep{Belokurov14}. In this plane, the Sgr-stream stars form a narrow sequence in each hemisphere, with velocities distinct from the MW. We use these sequences to refine our sample by selecting only stars broadly consistent with the Sgr signal. Figure~\ref{sgr_selection}a shows the position of stars from our sample in the $\Lambda$$_{\odot}$ vs V$_{GSR}$ plane, compared to the distribution of Sgr stars from~\citet{Belokurov14}. The Sgr stream is clearly visible in both hemispheres, with velocities different from the MW halo across most of the $\Lambda$$_{\odot}$ range. \\
To further decontaminate our sample we determine the distances to individual stars. Distances are obtained by selecting isochrone points with the correct g$-$i colour, and shifting their i-band magnitude to match the observations. We make use of isochrones from the Teramo/BaSTI library~\citep{TeramoI}, and adopt the age-metallicity relation of Sgr globular clusters to determine the correct age at each metallicity~\citep{Forbes10}. Subsequently, we pare our sample by keeping only those stars with distances consistent with the literature distance determinations to different branches of the stream from different distance indicators~\citep{Niederste-Ostholt10,Koposov12}.  \\
Figure~\ref{sgr_selection}b shows the photometric, MW-corrected i,g$-$i Colour-Magnitude Diagram~(CMD) of Sgr constructed from SDSS DR9 photometry, overlaid with the stars in our spectroscopic sample. For both photometric and spectroscopic points, the magnitudes are corrected for the distance gradient along the stream using literature distance determinations. The colours of the spectroscopic points indicate the spectroscopic [Fe/H] abundance. Isochrones representative of Sgr populations~(with ages from~\citet{Forbes10}) are also shown for comparison. \\ 
The stars from our sample form a clear Red Giant Branch~(RGB) with a metallicity trend consistent with evolution of stellar populations in a relatively isolated system. The RGB forms a continuation of the Sub-Giant and RGB branches seen in the photometric CMD, indicating that our sample indeed consists of likely Sgr members. Despite our selection of stars being consistent with Sgr distances, there may still be residual contamination by metal-poor MW halo stars overlapping with the blue, metal-poor RGB. To remove this contamination from our sample, we reject all blue stars with g$-$i$\le$1.0, to obtain the cleanest possible Sgr selection. \\
Our final sample consists of 313 stars with velocities and spatial positions consistent with the Sgr stream, which together form an RGB characteristic of a relatively isolated stellar system at the distance of Sgr. Figure~\ref{sgr_alphafe}a presents the $\alpha$-element distribution of our final Sgr stream sample. The [$\alpha$/Fe] ratios of the Sgr main body~(small open circles) are also shown for comparison, using the same weighting of individual $\alpha$-elements as adopted for SDSS/SEGUE estimates~\citep{SmeckherHane02,Carretta102,McWilliam13}. Furthermore, the median [$\alpha$/Fe] ratios of the MW components~(see Figure~\ref{MWcomparison}) are shown as the blue, dashed lines. Our sample of Sgr stars displays $\alpha$-element abundances that are in good agreement with results from HR studies for metal-poor stars~([Fe/H]$\le$$-$1.0). The abundance distribution of more metal-rich stars is not well traced in our sample, due to lack of metal-rich Sgr stars outside of the main body. However, the [$\alpha$/Fe] ratio of detect stars traces the upper envelope of the HR results and is consistent to within the error-bars. Furthermore, the $\alpha$-element abundances of the SDSS/SEGUE stars are also consistent with values derived for globular clusters associated to the Sgr stream~(large open circles in Figure~\ref{sgr_alphafe}a), giving further confidence to our results~\citep{Brown99,Carraro04,Cohen04,Tautvaisiene04,Mottini08}. \\
Figure~\ref{sgr_alphafe}a shows that the stars in our high-confidence Sgr sample cover a larger range of metallicities than the HR Sgr results, as a result of the larger fraction of metal-poor stars in the stream. The $\alpha$-element abundances of Sgr-stream stars are consistent with a flat plateau for metal-poor~([Fe/H]$\le$$-$1.5) stars. For more metal-rich stars, the $\alpha$-element abundances in Sgr remain flat for metallicities up to [Fe/H]$\approx$$-$1.3, below which they start to turn down. This is consistent with the presence of an $\alpha$-element ''knee" as a result of increased metal contributions to the ISM from SNe Ia explosions~\citep[e.g.,][]{Tinsley79}. This is the first time that the $\alpha$-element knee of the Sgr dSph has been robustly determined across a large range of metallicity.  \\
We determine the location of the knee by fitting a toy model consisting of a plateau at low metallicities followed by a linear decrease of [$\alpha$/Fe]. The best-fit toy model is shown in Figure~\ref{sgr_alphafe}a as the solid red line, along with 1$\sigma$ uncertainties~(grey shaded region). The best fit places the plateau at [$\alpha$/Fe]=0. 33, the knee at [Fe/H]=$-$1.27$\pm$0.05 and the $\alpha$-element slope at $-$0.16$\pm$0.02. Furthermore, if we fit the SDSS/SEGUE Sgr sample together with the HR main-body samples~(yellow solid line in Figure~\ref{sgr_alphafe}a), the knee is placed at [Fe/H]=$-$1.25$\pm$0.02, followed by a steeper $\alpha$-element slope of $-$0.26$\pm$0.01.

\section{Conclusions and discussion}
\label{conclusions}
We have used spectroscopic abundance results for stars from the SDSS/SEGUE survey to test the reliability of medium-resolution $\alpha$-element abundance determinations, and to study the $\alpha$-element distribution of the Sgr stellar stream. By selecting stars from different components of the MW, we have shown that SDSS/SEGUE mean [$\alpha$/Fe] ratios and dispersions are in good agreement with HR MW spectroscopic studies for our sample of S/N$\ge$50 stars~(Figure~\ref{MWcomparison}). \\
Using selection criteria based on the spatial position, radial velocity, distance and CMD colours we obtain a robust sample of Sgr-stream stars, mostly free from MW contaminants. When plotted in a CMD, our sample forms a clear RGB branch~(see Figure~\ref{sgr_selection}b), consistent with the distance and stellar content of Sgr. This shows that the final sample consists of bonafide Sgr stream members and is not strongly influenced by MW contamination. The low MW contamination is especially important, since the Sgr $\alpha$-element distribution closely resembles that of the MW. \\
Figure~\ref{sgr_alphafe}a shows the $\alpha$-element distribution of the Sgr stream. A clear turn-down in $\alpha$-element abundances is visible at [Fe/H]$\approx$$-$1.3, which is consistent with the presence of a ''knee" due to metal contributions from SNe Ia explosions~\citep{Tinsley79}. We note, however, that other authors have argued that this feature can also be created by inferring a top-light IMF for Sgr~\citep{McWilliam13}. Fitting a toy model to our data we determine, for the first time, the position of the knee in the Sgr dSph, which takes place at [Fe/H]$\approx$$-$1.3. Comparison to the age-metallicity relation from~\citet{Forbes10} shows that the the knee corresponds to an age of $\approx$11 Gyr, roughly 1.5 Gyr after their proposed start of star formation in Sgr. This time delay between the onset of SN Ia is consistent with theories of galaxy formation, which place the occurrence of the knee approximately 1 Gyr after the start of star formation~\citep[see,~e.g.][]{Raiteri96,Matteucci01}. \\
Comparison of our medium-resolution $\alpha$-element abundances to results from HR studies~(black, open circles in Figure~\ref{sgr_alphafe}) show a partial overlap in metallicities between both samples. However, the majority of the HR sample displays [$\alpha$/Fe] ratios well below those of the Sgr stream, and at higher metallicity. This could be due to the increased and prolonged star formation in the Sgr core compared to the stream~(such as induced star formation by the accretion event), which was likely stripped from the outskirts of the Sgr dSph. This increased star formation may have led to a more rapid decline in [$\alpha$/Fe] ratios in the main body that is not reflected in the stream. However, for a complete analysis of the enrichment differences between the stream and main body, we will need to compare ages of stars in both components to the estimated time of initial accretion of Sgr. For this, the detailed star formation history of the Sgr stream needs to be determined. \\
Figure~\ref{sgr_alphafe}b shows the $\alpha$-element abundance trends of Sgr-stream stars in comparison to the MW~(small grey dots) and other LG dSphs with HR $\alpha$-element measurements~\citep{Venn04, Pompeia08,Tolstoy09, Letarte10,Lemasle12,Lapenna12,Vanderswaelmen13}. Solid coloured lines show the best-fit toy model fit to the distribution for each dSph, assuming the level for the $\alpha$-plateau is the same as for the Sgr stream. Note that in the case of Carina, we have only fit the model to the stars associated to the oldest star formation episode~\citep[][de Boer et al, in prep]{Lemasle12}. For systems without a robust knee detection~(Fornax and LMC), the lines trace the outer envelope of the $\alpha$-element abundance distribution. \\
Comparison to the MW distribution shows that the $\alpha$-knee of the Sgr stream is only slightly less metal-poor than the $\alpha$-knee in the more massive MW, which is fit at [Fe/H]=$-$1.04$\pm$0.02. The position of the $\alpha$-knee in other LG dSph galaxies is also correlated with the mass of the system. The knee in the small Carina dSph~(M$_{dyn}$($\le$r$_{h}$)$\approx$3.4$\times$10$^{6}$ M$_{\odot}$) takes place at [Fe/H]=$-$2.72$\pm$0.27, while the knee in the more massive Sculptor dSph~(M$_{dyn}$($\le$r$_{h}$)=1.4$\times$10$^{7}$ M$_{\odot}$) is fit at [Fe/H]=$-$1.87$\pm$0.09~\citep{Walker11,McConnachie12}. \\ 
The $\alpha$-knee of the massive Fornax dSph has not been robustly determined from HR spectroscopy~\citep{Letarte10}. However, the $\alpha$-abundance distribution of the metal-rich stars in Fornax suggests that the metallicity of the $\alpha$-knee is in between the position of the Sculptor dSph and the position determined for the Sgr knee in this work~\citep{deBoer2012B}. This would be consistent with the mass of Fornax~(M$_{dyn}$($\le$r$_{h}$)$\approx$1.6$\times$10$^{8}$ M$_{\odot}$) which is in between Sculptor and Sgr~\citep{Walker06}. Furthermore, the knee of the even more massive LMC~(M$_{dyn}$=1.7$\times$10$^{9}$ M$_{\odot}$) is expected at metallicities comparable to Sgr or the MW~\citep{VanderMarel14}. Therefore, the position of the $\alpha$-knee determined in this work is consistent with the mass determined for Sgr. \\
In conclusion, Figure~\ref{sgr_alphafe}a shows that early accretion~(before the $\alpha$-knee formed) of Sgr-like galaxies could have contributed significantly to the build-up of the MW stellar halo system. In particular, the accretion of Sgr-like galaxies could be important for the formation of the metal-rich inner parts of the halo system, which are not easily assembled through the merging of small galaxies with an $\alpha$-knee at low metallicity~\citep[e.g.,][]{Tolstoy09}. In conclusion, determining the detailed properties of galaxies like Sgr can help us to further our understanding of the early formation and evolution of larger and more complex systems such as the MW and Andromeda.

\section*{Acknowledgements}
The research leading to these results has received funding from the European Research Council under the European UnionÕs Seventh Framework Programme (FP/2007-2013) / ERC Grant Agreement n. 308024. VB acknowledges financial support from the Royal Society. T.d.B. acknowledges financial support from the ERC. T.C.B. acknowledges partial support for this work from PHY 08-22648; Physics Frontier Center/Joint Institute or Nuclear Astrophysics (JINA), awarded by the US National Science Foundation. Y.S.L. is a Tombaugh Fellow at New Mexico State University. The authors would also like to thank the referee for his comments, which helped to improve the paper. \\
Funding for SDSS-III has been provided by the Alfred P. Sloan Foundation, the Participating Institutions, the National Science Foundation, and the U.S. Department of Energy Office of Science. The SDSS-III web site is http://www.sdss3.org/. \\
SDSS-III is managed by the Astrophysical Research Consortium for the Participating Institutions of the SDSS-III Collaboration including the University of Arizona, the Brazilian Participation Group, Brookhaven National Laboratory, Carnegie Mellon University, University of Florida, the French Participation Group, the German Participation Group, Harvard University, the Instituto de Astrofisica de Canarias, the Michigan State/Notre Dame/JINA Participation Group, Johns Hopkins University, Lawrence Berkeley National Laboratory, Max Planck Institute for Astrophysics, Max Planck Institute for Extraterrestrial Physics, New Mexico State University, New York University, Ohio State University, Pennsylvania State University, University of Portsmouth, Princeton University, the Spanish Participation Group, University of Tokyo, University of Utah, Vanderbilt University, University of Virginia, University of Washington, and Yale University. 
\bibliographystyle{mn2e}
\bibliography{Bibliography}

\label{lastpage}

\end{document}